# Amplifying spin waves along Néel domain wall by spin-orbit torque

Xiangjun Xing,[1,a)] T. Wang,[1] and Yan Zhou[2,a)]

[1]*School of Physics & Optoelectronic Engineering, Guangdong University of Technology, Guangzhou 510006, China*

[2]*School of Science & Engineering, The Chinese University of Hong Kong, Shenzhen, Guangdong 518172, China*

**Abstract**

Traveling spin waves in magnonic waveguides undergo severe attenuation, which tends to result in a finite propagation length of spin waves, even in magnetic materials with the accessible lowest damping constant, heavily restricting the development of magnonic devices. Compared with the spin waves in traditional waveguides, propagating spin waves along strip domain wall are expected to exhibit enhanced transmission. Here, we demonstrate, theoretically and through micromagnetic simulations, that spin-orbit torque associated with a ferromagnet/heavy metal bilayer can efficiently control the attenuation of spin waves along a Néel-type strip domain wall, despite the complexity in the ground-state magnetization configuration. The direction of the electric current applied to the heavy-metal layer determines whether these spin waves are amplified or further attenuated otherwise. Remarkably, our simulations reveal that the effective current densities required to efficiently tune the decay of such spin waves are just ~$10^{10}$ Am$^{-2}$, roughly an order smaller than those required in conventional spin waveguides. Our results will enrich the toolset for magnonic technologies.

**Keywords:** spin wave, strip domain wall, attenuation, spin-Hall effect

---

[a)]Authors to whom correspondence should be addressed: `xjxing@gdut.edu.cn` and `zhouyan@cuhk.edu.cn`

Propagating spin waves dissipate energy because of scattering and damping, resulting in gradually reduced amplitude of spin waves away from the source, which hampers widespread use of magnonic devices, although they are reckoned more competitive than CMOS-based electronic devices in non-Boolean logic and computation[1] and special-type data processing, e.g., image recognition.[2,3]

Recently, several sorts of magnonic waveguides[4–9] were suggested that exhibit enhanced transmission of spin waves by circumventing certain essential scattering channels. Amongst them the magnonic waveguide based on domain walls possesses substantial advantages, such as high energy efficiency, simple initialization procedure, and easy adaptation to sample geometry.[10] Noteworthily, some latest experimental studies[11,12] have demonstrated the guidance and steering of spin waves through domain walls embedded in extended geometries as straight, bent, or arbitrarily shaped waveguides. An especially important aspect lies in the realization of a converging domain walls configuration,[11] which can serve as fundamental building blocks for constructing reconfigurable domain wall-based magnonic circuitry.[13]

Despite suppressed scattering and elevated group velocity,[6–8] the decay of propagating spin waves along domain walls due to damping remains huge.[6] Accordingly, it is necessary to further diminish the damping-induced energy loss. In conventional magnonic waveguides, this can be attained by using the Zhang-Li or Slonczewski type spin torques as revealed in early theoretical studies.[14,15] Currently, the same task can be performed using spin-orbit torques, as shown theoretically and experimentally in Refs.[16–20]. However, although the principle of controlled damping is already known for more than a decade,[21] transplanting such implementations to domain wall-based magnonic waveguides is not direct, since the environment met herein by such spin waves becomes very complex; the background magnetization configuration is no longer a quasiuniform single domain.

In the Letter, we address this issue analytically and examine the theoretical finding using micromagnetic simulations. Our theory shows that the spin-wave attenuation in domain wall-based magnonic waveguides can be controllably manipulated by the Slonczewski spin torque, which arises from the spin-Hall effect in a ferromagnet/heavy metal bilayer. The simulations qualitatively confirm the key aspect of theoretical result but predict smaller current densities effective for controlling the spin-wave attenuation. These results offer useful insights for developing relevant magnonic devices.

We consider a narrow magnetic nanowire [Fig. 1(a)] that can be lithographically patterned from an ultrathin ferromagnet/heavy metal (FM/HM) bilayer,[22–24] in which the magnetic free energy includes Heisenberg exchange interaction ($A$ denotes the exchange stiffness), interfacial Dzyaloshinskii-Moriya interaction (DMI, $D$ is the DMI strength), perpendicular magnetocrystalline anisotropy (PMA, $K_u$ is the PMA constant), and magnetostatic interaction ($M_s$ denotes the saturation magnetization). When an electric current is injected into the structure, spin orbit torques (denoted as $\mathbf{T}_S$) will be imposed on the magnetic moments in the FM layer owing to the spin-Hall effect.[25] The extended Landau-Lifshitz-Gilbert equation with the spin orbit torques is used to track the current-induced dynamics of spin waves in the nanowire that harbors a longitudinal magnetic domain wall,

$$\frac{\partial \mathbf{m}}{\partial t} = -\gamma \mathbf{m} \times \mathbf{H}_{eff} + \alpha \mathbf{m} \times \frac{\partial \mathbf{m}}{\partial t} + \mathbf{T}_S, \quad (1)$$

where $\mathbf{m} = \mathbf{M}/M_s$ with the magnetization $\mathbf{M} = \mathbf{M}(x, y, t)$ varying with the spatial coordinates $x$ and $y$ as well as the time $t$, and $\mathbf{H}_{eff}$ is the effective field; $\gamma$ and $\alpha$ are the gyromagnetic ratio and Gilbert damping constant, respectively.

For this system, the magnetostatic interaction is absorbed into the uniaxial anisotropy to simplify the theoretical calculation, and thus the effective field reads,

$$\mathbf{H}_{eff} = \mathcal{A}\nabla^2 \mathbf{m} - \mathcal{D}\left[\frac{\partial m_z}{\partial x}\hat{x} + \frac{\partial m_z}{\partial y}\hat{y} - \left(\frac{\partial m_x}{\partial x} + \frac{\partial m_y}{\partial y}\right)\hat{z}\right] + \mathcal{K}m_z\hat{z}, \quad (2)$$

where $\mathcal{A} = \frac{2A}{\mu_0 M_s}$, $\mathcal{D} = \frac{2D}{\mu_0 M_s}$, and $\mathcal{K} = \frac{2K}{\mu_0 M_s}$ with $\mu_0$ denoting the vacuum permeability and $K = K_u - \frac{1}{2}\mu_0 M_s^2$.[26] Regarding the spin orbit torques, we take into account only the damping-like term and ignore the field-like one (supplementary material Fig. S1), namely,

$$\mathbf{T}_S = -\gamma a_J \mathbf{m} \times (\mathbf{m} \times \boldsymbol{\sigma}), \quad (3)$$

where the electrons' spin polarization direction $\boldsymbol{\sigma} = \hat{\mathbf{J}} \times \hat{\mathbf{z}}$ with $\hat{\mathbf{J}}$ denoting the unit vector along the electrical current [Fig. 1(a)], and $a_J = \frac{\hbar J G \Phi_H}{e \mu_0 M_s d_{FM}}$ with $\hbar$ denoting the reduced Planck constant, $e$ the elementary charge, $\Phi_H$ the spin Hall angle, $J$ the electrical current density, and $G = \frac{\Lambda^2}{(\Lambda^2+1)+(\Lambda^2-1)(\mathbf{m}\cdot\boldsymbol{\sigma})}$ measuring the spin torque asymmetry determined by the relative orientation of $\mathbf{m}$ and $\boldsymbol{\sigma}$ in the FM layer. For simplicity, we take $\Lambda = 1$, i.e., $G = 1/2$, as in most literature.[20,27]

In this study, we are interested in small-amplitude spin waves, for which the linear approximation is justified, that is, the variable magnetization can be expanded in a series of plane waves of magnetization.[14,15] Consequently, $\mathbf{m}$ can be written as follows,

$$\mathbf{m}(x,y,t) = \mathbf{m}^g(y) + \mathbf{m}^0(y) e^{-x/L} e^{i(\omega t - kx)}, \quad (4)$$

where $\omega$, $k$, and $L$ are the angular frequency, wavenumber, and attenuation length of spin waves, respectively. The ground-state magnetization configuration $\mathbf{m}^g(y)$, i.e., the vertically oriented opposite domains separated by a longitudinal domain wall, is mimicked using the following form (supplemental material Fig. S2), $\mathbf{m}^g(y) = \left(0, \text{Sech}\left(\frac{y}{\Delta}\right), \text{Tanh}\left(\frac{y}{\Delta}\right)\right)$, where $\Delta = \sqrt{A/K}$ is the domain-wall thickness. The excitation amplitude of the spin waves, $\mathbf{m}^0(y)$, is inhomogeneous across the width of the nanowire;[7] we assume that $\mathbf{m}^0(y) = (m_x^0, m_y^0, m_z^0) \text{Sech}\left(\frac{y}{\Delta}\right)$ where $m_y^0 \ll m_x^0, m_z^0$.

Substituting Eqs. (2)–(4) into Eq. (1), then letting $y = 0$, and denoting $\Psi = e^{-x/L} e^{i(\omega t - kx)}$ and $\gamma_0 = \frac{\gamma}{1+\alpha^2}$, one obtains supplementary material Eqs. (S1). Ignoring the high-order terms of $\Psi$ in Eqs. (S1) and after some algebra, one will get two complex equations with respect to $m_x^0$ and $m_z^0$, as shown in supplementary material Eqs. (S2). Combining Eqs. S2a and S2b, eliminating $m_x^0$ and $m_z^0$, neglecting the small terms proportional to $\alpha^2$, and keeping in mind that $\text{Re}[\omega] = \omega_f$ and $\text{Im}[\omega] = 0$ as well as $1/L^2 \ll k^2$, one can arrive at,

$$L = \frac{\mathcal{D}\omega_f\Delta + \gamma_0 k[\Delta(\mathcal{D}^2 - 2\mathcal{A}^2 k^2 + 2\mathcal{A}\mathcal{K}) - 2\mathcal{A}\mathcal{D}]}{-\alpha\omega_f(\mathcal{D} + \mathcal{A}k^2\Delta - \mathcal{K}\Delta) + a_J\Delta(\omega_f + \gamma_0\mathcal{D}k)}. \tag{5}$$

Eq. (5) appears a bit complicated, but it is consistent, in spirit, with the counterparts in Refs.[14,15,20]. Here, the complexness of Eq. (5) originates from the complexity of the studied system, which includes more complex magnetization background as well as two extra interactions, namely, the PMA and interfacial DMI. Accordingly, the related parameters $\Delta$, $K$, and $D$ enter and introduce several additional terms in the equation. Despite the seeming complexness in form, Eq. (5) gives a similar $L(J)$ dependency [Fig. 1(b)] as in Refs.[14,15].

Choosing the typical experimental values reported for the HM/FM bilayer systems:[22–24] $A = 15$ pJm$^{-1}$, $D = 3.5$ mJm$^{-2}$, $K_u = 0.8$ MJm$^{-3}$, and $M_s = 580$ kAm$^{-1}$ and supposing that $\Phi_H = 0.13$, $\alpha = 0.015$,[28,29] and $d_{\text{FM}} = 1$ nm, we plot $L(J)$ in Fig. 1(b) according to Eq. (5). Clearly, the curves for different frequencies display the same tendency, that is, the spin-wave attenuation length increases with the increased current density before a threshold current density, and becomes negative at the threshold current density. This suggests that for the positive currents along $+x$, the spin-wave attenuation is suppressed by the current via the spin-Hall effect; by contrast for the negative currents, i.e., along $-x$, the spin waves decay faster. At the threshold current density, the Gilbert damping torque is completely compensated by the current-induced Slonczewski toque; consequently, the spin waves propagate with constant amplitudes, giving rise to an infinite attenuation length. The threshold current densities change with the frequencies; spin waves at lower frequencies require smaller currents to reach the equiamplitude propagation state.

For the suprathreshold current density, the attenuation length of spin waves turns negative, where the spin waves propagate without decay in amplitude, and instead, they exhibit ever larger amplitude along the propagation path. However, in this situation, the spin waves will become instable after traveling a certain distance, because of chaotic dynamics associated with the strong nonlinear effect at large-angle magnetization precession around the effective field. These behaviors have been seen in Refs.[14,15,18,20]. High current densities induce chaotic dynamics and even worse may directly damage the device itself through producing excessive Joule heating. For the spin waves along a Néel wall, there exists an upper threshold for the current density (~ $5.0\times10^{12}$ A/m$^2$), above which the domain-wall structure will be destroyed by the spin-orbit torque, as clarified in our previous investigations[10,30]. Noteworthily, in the present study, the effective current densities for amplifying the spin waves are only ~ $10^{10}$–$10^{11}$ A/m$^2$. For the stated reasons, the plot in Fig. 1(b) cannot extend to $J \to \infty$, where the result is meaningless.

Overall, the theoretical effective working current densities on the order of $10^{11}$ Am$^{-2}$ are consistent with those of spin waves in conventional waveguides hosting a quasiuniform magnetic background, as identified experimentally[16–19] and via micromagnetic simulations.[15,20]

As a further step, we resort to micromagnetic simulations to check the theoretical results and uncover the full details of the dynamic process by numerically solving Eq. (1), in which the magnetostatic interaction is incorporated without any truncation and the spin-torque asymmetry is involved by setting $\Lambda = 2$.

The finite-difference code, OOMMF,[31] is used to perform the numerical calculations, in which we consider only the FM layer explicitly and the effects arising from the HM layer is absorbed into the parameters. The nanowire has a width of 60 nm and a length of 2 μm. Firstly, the nanowire is initialized in the equilibrium magnetization configuration containing a strip domain wall, and then the spin-wave dynamics in the nanowire is surveyed following an applied current. The spin waves are excited by a local alternating field, which can be created experimentally using a narrow antenna, as shown in Fig. 1(a). Here, the same material parameters are used as in the analytical calculation. For computation, we discretize the FM layer into an array of $1 \times 1 \times 1$ nm$^3$ boxes, whose side length is much smaller than the exchange length $l_{ex} = \sqrt{2A/\mu_0 M_s^2} \approx 8.4$ nm. We introduce an absorbing boundary condition[32,33] to suppress the reflection of spin waves near the terminals.[7,34] The results from the simulations are presented in Figs. 2–4.

Figure 2(a) displays the frequency spectra of the spin waveguides responding to applied electric currents. Apparently, these spectra are strongly modulated by the currents at least in the indicated frequency range. Oscillatory spectra of the amplitude versus frequency are a generic feature and reflect the intrinsic nature of spin-wave excitations in confined magnetic samples[33,35–37]. Each peak in the spectra rises with the increased current density. This is a clear evidence that the spin-orbit torque is capable of amplifying the spin waves in domain wall-based waveguides, despite the highly inhomogeneous magnetization distribution. For the spin waves below 13 GHz, the current's modulation efficiency is particularly high. Thus, we derive the attenuation lengths of spin waves at several separate frequencies in this range and plot them in Fig. 2(b) against the current density. The evolution of $L$ versus $J$ clearly indicates that the spin waves are well tuned by the applied current. While positive currents amplify the spin waves, negative currents further attenuate them; experimentally, such situations can be realized readily by reversing the direction of the current in the heavy metal layer. Without applied current, i.e., $J = 0$, the spin waves at different frequencies exhibit similar attenuation lengths, $L \sim 1100$ nm, only with slight spreading. As $J$ increases, the attenuation lengths at individual frequencies diffuse and appear to exhibit a larger interval, implying that the amplification efficiency depends on the frequency, which is in agreement with the theoretical result, i.e. Eq. (5) as $L$ is the function of $J$ and $\omega_f$.

However, two striking differences can be found by comparing the simulated and analytical $L(J)$ dependences. First, the effective range of the working current densities differs by an order of magnitude. While the theoretical threshold current density is around $2.0\times10^{11}$ Am$^{-2}$, the simulated one is down to $3.0\times10^{10}$ Am$^{-2}$. Second, the theoretical values of the attenuation lengths are much larger than the simulated ones. At $J = 0$, the theoretical values of $L$ are above 2 μm, roughly twice the simulated values ~1 μm. These discrepancies between the theory and simulations may be due to incomplete inclusion of the magnetostatic interaction in the effective field, ignorance of the spin torque asymmetry, and inaccurate description of the real profile of the complex magnetization background in the theory. We find that varying the forms of $\mathbf{m}^g(y)$ will yield distinct threshold current densities when all other factors are being fixed (an example is shown in supplementary material Fig.S2). Undoubtedly, the simulation results better reflect the reality, whereas the theory simply captures the core of the question.

Now, let us get into the 2D spatial distribution of the spin waves in domain wall-based waveguides under various current densities, which directly illustrate the modulating effect of currents on the spin waves. We choose typical frequencies 4.1, 5.3, and 16.6 GHz from three separate regions in Fig. 2(a) with distinct spectral characters and present their modal profiles in Figs. 3(a)–(e) and supplementary material Figs. S3 and S4. The spatial oscillation in each panel of Figs. 3(a)–(e) is due to interference of the propagating spin waves and the partly reflected ones from the edge of the absorbing boundary [i.e., the shaded area on the waveguide in Fig. 1(a)], featuring a beating pattern with spatially modulated amplitude[34,38,39]. As shown in Figs. 3(a)–(c), under an applied zero or negative current, the spin waves become dimmer with the propagation distance. However, if a positive current is applied, the spin waves will travel farther or, in other words, exhibit enhanced oscillating strength [Fig. 3(d)]. Once the current density surpasses the threshold, the spin waves will be overamplified, i.e., their amplitudes increase with the propagation distance. In this situation, the attenuation lengths turn to be negative [Fig. 3(e)]. Figs. 3(f)–(j) show the steady-state propagation patterns of spin waves at a certain time. As the counterparts of Figs. 3(a)–(e), these snapshots contain both the amplitude and phase information and allow one to identify a feature typical of the wave beams under current, that is, as the current density deviates from zero, the crests and troughs of the spin waves are not collinear with the domain wall but separated in the transverse direction. Opposite currents lead to reversed displacement between the wave crests and troughs. In Fig. 4, we clarify the physics for spin-wave amplification as well as crests and troughs' displacement induced by spin-orbit torque according to a unified picture, in which the magnetic moments in the domain wall and in the domains exhibit various dynamics under the combined action of the spin-orbit torque and effective field.

Finally, we extract the dispersion relations of the spin waves under different current densities, as shown in Figs. 5(a)–(e), a comparison of which can answer whether the Doppler effect occurs to such spin waves in the considered current range. The fact that, no considerable shift along the frequency or wavenumber axis can be found among the five sets of curves, implies that the spin-wave Doppler effect is trivial in the covered range of current density, although remarkable amplification of spin waves is enabled in the same range. This result is reasonable because in conventional waveguides Doppler shifts will be appreciable only if the current densities exceed $1.0 \times 10^{12}$ Am$^{-2}$.[15]

Spin waves along the domain wall (bound mode) possess a lower frequency gap than those outside the domain wall (bulk mode).[6–8] Wang and Wang derived explicitly the dispersion relations and group velocities for the bound and bulk modes, clearly indicating that at an identical frequency the bound mode travels a longer distance than the bulk mode.[8] Accordingly, the bound mode has lower attenuation compared to the bulk mode. However, as shown in Ref.[6], the bound mode remains to experience considerable decay because of the Gilbert damping.

Refs.[14–20] represent a full spectrum of current-induced control of spin waves. A decade ago, Seo *et al.*[14] and Xing *et al.*[15] considered, theoretically and through micromagnetic simulations, the effects of Zhang-Li and Slonzewski type spin torques, respectively, on backward-volume spin waves in a Py magnetic nanostrip, where the static magnetization is directed along the strip length. Later on, the control of propagating spin waves in CoFeB/Ta, CoFeB/Pt, YIG/Pt, and Py/Pt bilayers were experimentally implemented in Refs.[16–19], respectively, by exploiting the spin-Hall based spin-orbit torque. Recently, Woo and Beach[20] addressed micromagnetically and analytically the modification of propagating spin-wave attenuation in a Py/Pt bilayer based on spin-Hall effect. While those theories, simulations, and experiments[14–20] considered various kinds of spin torques and/or materials, they all adopted uniform magnetization as the static magnetic background to transport spin waves.

The present study demonstrates controllable amplification or attenuation of spin waves along Néel domain walls by spin-Hall based spin-orbit torque. Apart from the difference in the static magnetization background between our model (complex nonuniform magnetization) and those in Refs.[14–20] (simple uniform magnetization), we identify emerging dynamics of spin waves under spin-orbit torque [Figs. 3(f)–3(j)] and clarify the underlying mechanism (Fig. 4). The observed dynamics and associated physics have not been reported yet.

In conclusion, the effect of electric current on the spin waves in domain wall-based waveguides is examined theoretically and via micromagnetic simulations. Both the theory and simulations suggest that current can amplify or further attenuate such spin waves (depending on the current direction) by virtue of the spin-orbit torque, albeit the complexity of the dynamics in this system. Different from the counterparts in magnonic waveguides with a uniform magnetization configuration,[14–20] the spin waves propagating along domain walls exhibit displaced crests and troughs when the current is applied. The working current density on the order of $10^{10}$ Am$^{-2}$ for spin-wave amplification is very appealing and useful in practical applications of the domain wall-based magnonic devices.

See the supplementary material for those intermediate complex equations leading to the $L(J)$ formula, as well as the effect of the field-like torque on spin-wave amplification relative to the damping-like torque, the dependence of the theoretical $L(J)$ curve on the representation of $\mathbf{m}^g(y)$, and the mode profiles of spin waves at 5.3 and 16.6 GHz for various current densities.

X.J.X. acknowledges the support from National Natural Science Foundation of China (Grant No. 11774069). Y.Z. acknowledges the support by Guangdong Special Support Project (2019BT02X030), Shenzhen Peacock Group Plan (Grant No. KQTD20180413181702403), Pearl River Recruitment Program of Talents (2017GC010293), and National Natural Science Foundation of China (Grants No. 11974298 and No. 61961136006). Author contributions: X.J.X. conceived and initiated the study. All authors contributed to the analysis of the results. X.J.X. and Y.Z. wrote the manuscript.

**DATA AVAILABILITY**

The data that support the findings of this study are available from the corresponding author upon reasonable request.

**FIGURE CAPTIONS**

**Fig. 1** (Color online) Tuning of the propagating spin waves in domain wall-based magnonic waveguides by spin-Hall effect. (a) Setup of the study. In the multilayer, we assume that the FM layer is on top of the HM layer (not shown). The white dashed line along the *x*-axis of the nanowire marks a Néel domain wall, in which the magnetic moments point in the *y*-axis, as denoted by the black arrows. The symbols (⊙ up, ⊗ down) indicate the magnetization orientations in the opposite magnetic domains separated by the domain wall. For a positive current as marked by the purple arrows, the electrons' spin orientation, **σ** = **Ĵ** × **ẑ**, is along -*y*, as indicated by the grey arrows. The orange nanostrip near the left end of the nanowire is the excitation antenna, where an alternating field is exerted to emit spin waves. In the shaded areas near the ends, much higher damping coefficients are assumed in order to suppress spin-wave reflection. As an illustrative example, a snapshot of propagating spin wave along the domain wall is inserted for a sample 60 nm in width and 2 μm in length. (b) Theoretic spin-wave attenuation length against current density.

**Fig. 2** (Color online) Simulated current-controlled spin wave characteristics. (a) Spin-wave amplitude versus frequency at various current densities. The amplitude is averaged over entire sample volume. (b) Spin-wave attenuation length as function of current density.

**Fig. 3** (Color online) Spatial distribution of spin waves at different current densities. (a)–(e) Fourier amplitude distribution (viz. mode profile) of spin waves. The frequency is 4.1 GHz and the current density is indicated in each panel. The corresponding snapshots of spin waves in steady state are given in (f)–(j). As the current increases, the amplifying effect on the spin waves dominates. Here, the nanowire in each panel is 60 nm in width and 2 μm in length.

**Fig. 4** (Color online) Mechanism for amplification and the crests and troughs' shift of spin waves under spin-orbit torque. The electric current in the heavy-metal layer is along $x$ and $-x$ for panels (a)–(c) and (d)–(f), respectively. (a)(d) Effective field distribution, with yellow and green coding the up and down orientations of the effective field, respectively, and arrows repesenting its in-plane component. (b)(e) Precession of the $m_y$ component (blue arrows) in the magentic domains around the perpendicular effective field (⊙up, ⊗down) giving rise to a $m_x$ component (red arrows). The dashed green line denotes the domain-wall center. In panels (b) and (e), the electrons' spin orientation, $\boldsymbol{\sigma} = \hat{J} \times \hat{z}$, is along $-y$ and $y$, respectively, as indicated by the grey arrows. Consequently, in the two magentic domains, the spin-orbit torque acting on $m_z$, $\mathbf{T}_S \sim J[m_z\hat{z} \times (\boldsymbol{\sigma} \times m_z\hat{z})] \to \boldsymbol{\sigma}$, induces the small $m_y$ component, which then precesses around the perpendicular effective field, resluting in the small $m_x$ component. Under spin-wave excitation, magnetic moments (**m**) in the domain wall precess around the local effective field (**H**$_{eff}$) along $y$. Once a current is applied, these magnetic moments will be dragged (b) away from or (e) toward the local field by the spin-orbit torque (**T**$_S$) depending on the direction of $J$, leading to spin-wave amplification or suppression. (c)(f) Magnetization distribution with red and blue coding the right and left orientations of $m_x$, respectively, and arrows repesenting its in-plane component. The small static $m_x$ component superposes on the dynamic $m_x$ component of spin waves, engendering the noncollinear crests and troughs shifted in the transverse direction. Panels (a), (c), (d), and (f) are slices of a 60 nm wide and 2 μm long nanowire.

**Fig. 5** (Color online) Spin-wave dispersion characteristics at various current densities. (a) $J = -2\times10^{10}$ Am$^{-2}$, (b) $J = -1\times10^{10}$ Am$^{-2}$, (c) $J = 0$, (d) $J = 1\times10^{10}$ Am$^{-2}$, and (e) $J = 2\times10^{10}$ Am$^{-2}$. No obvious shift can be identified between these dispersion curves.

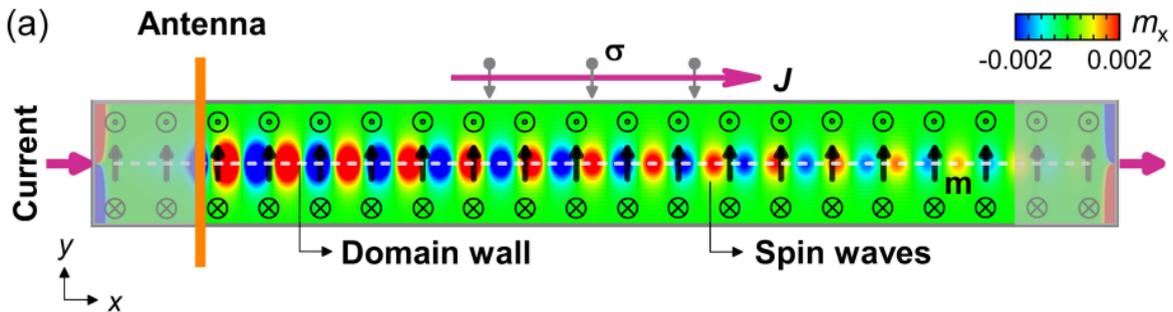

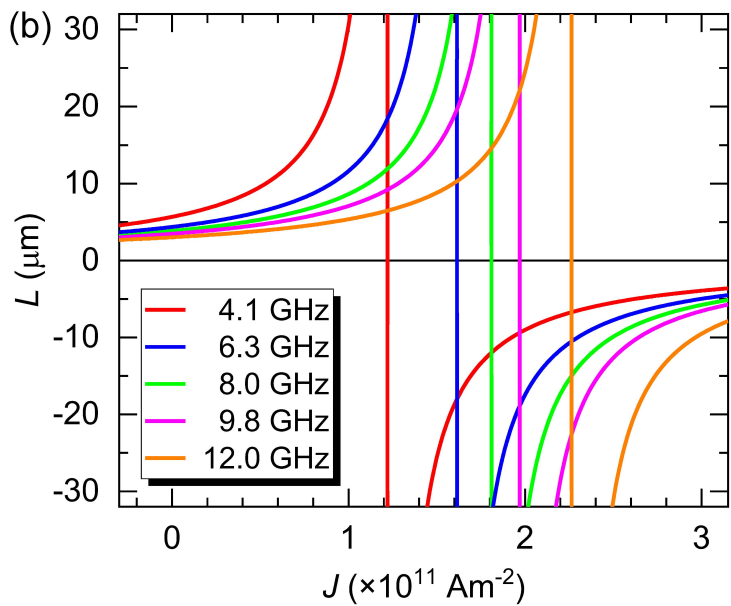

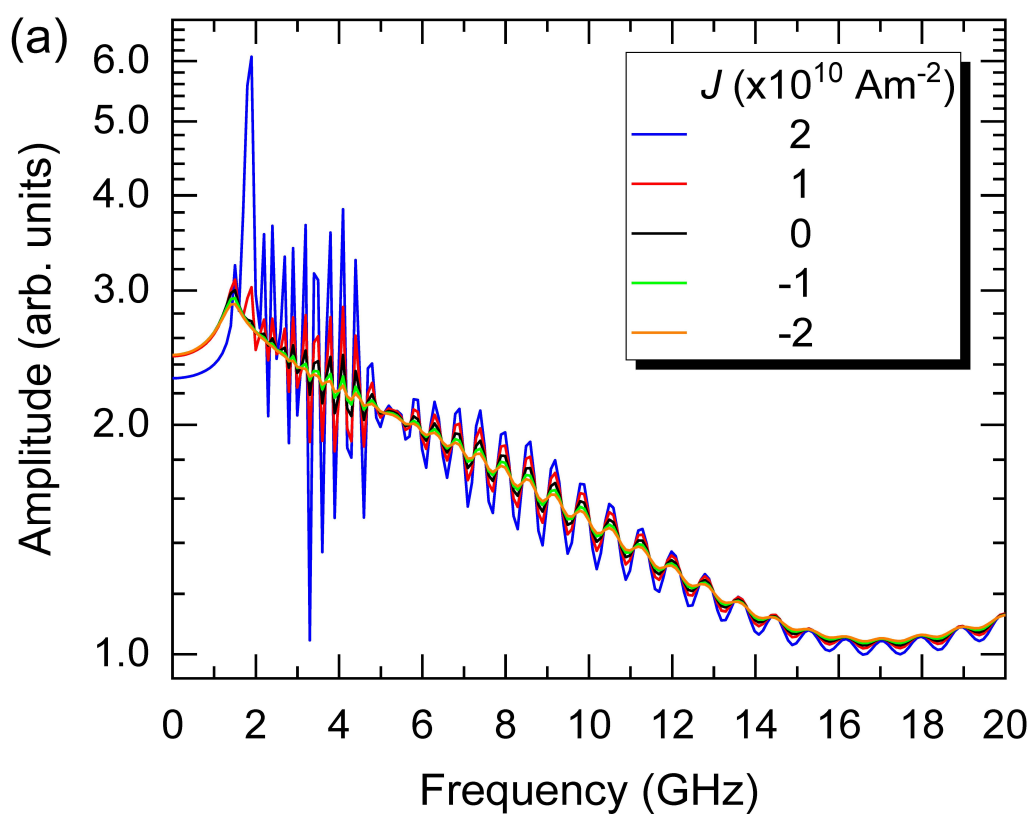

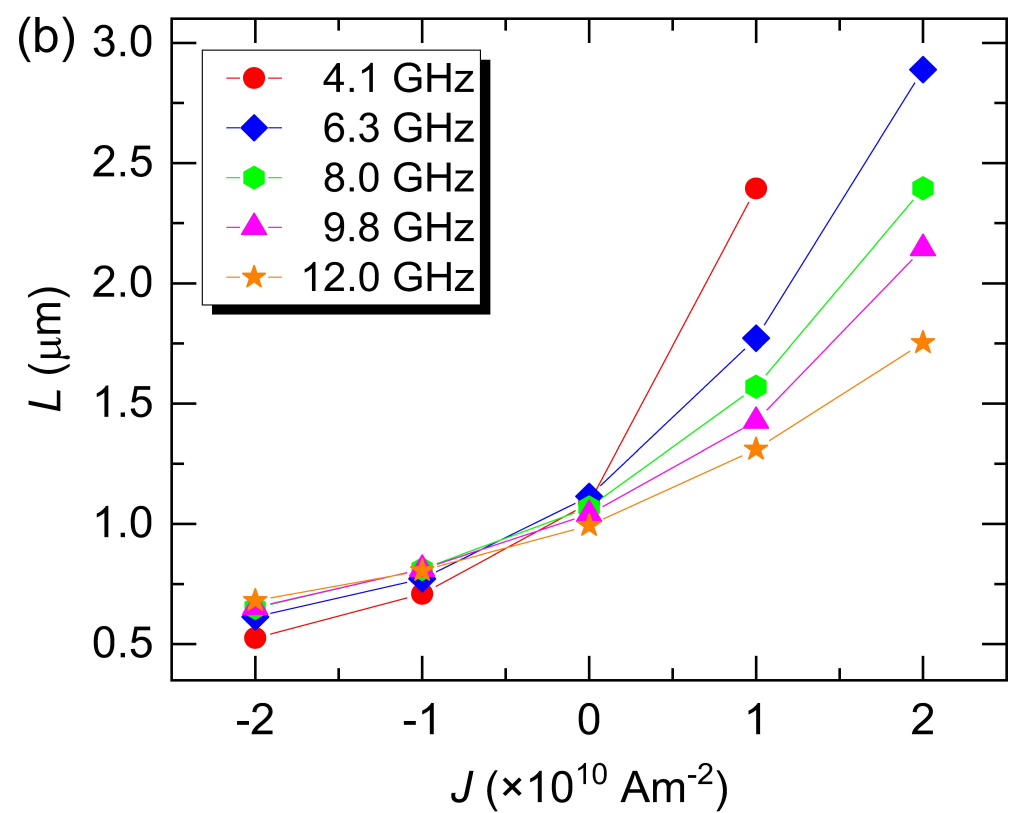

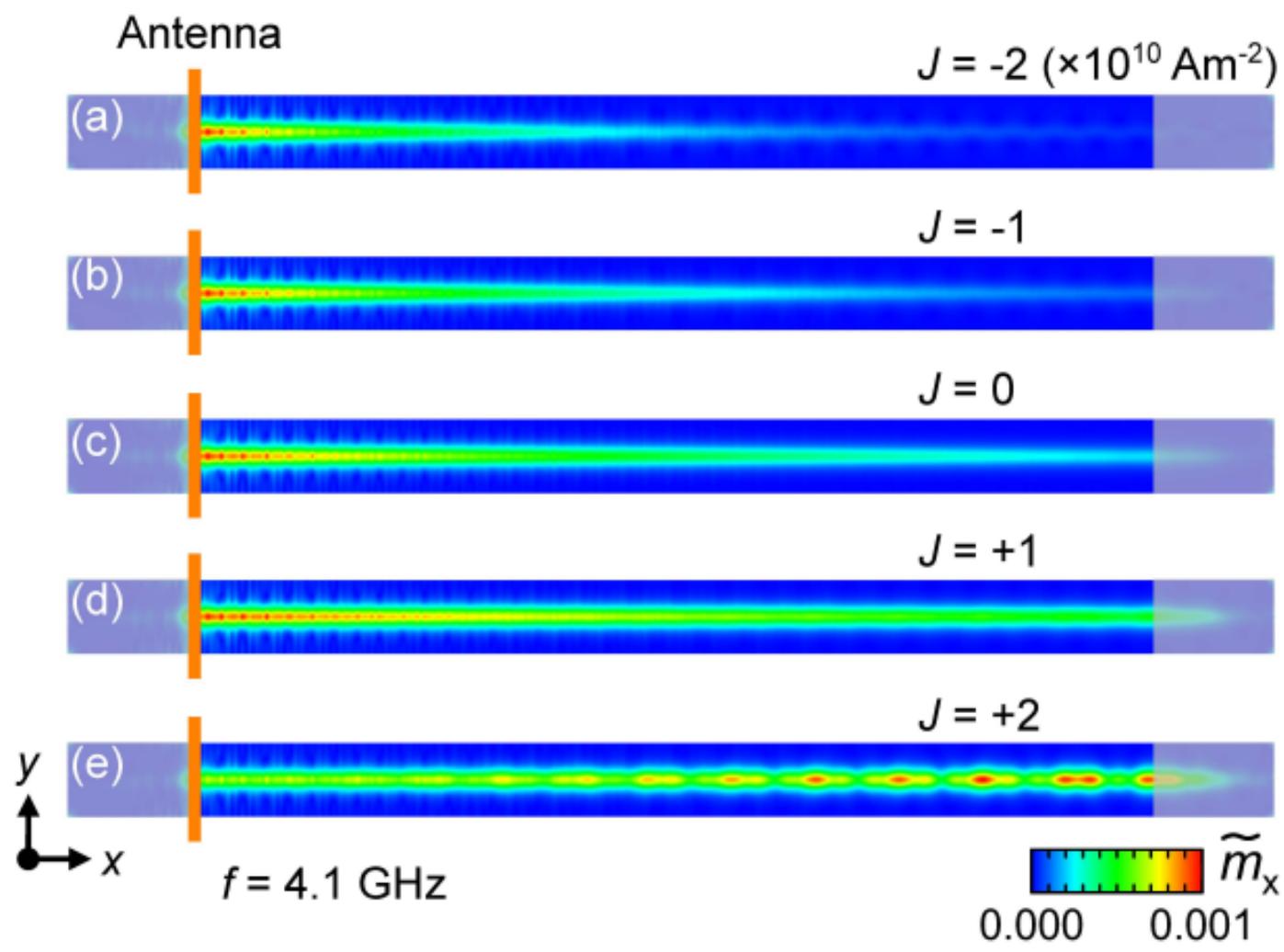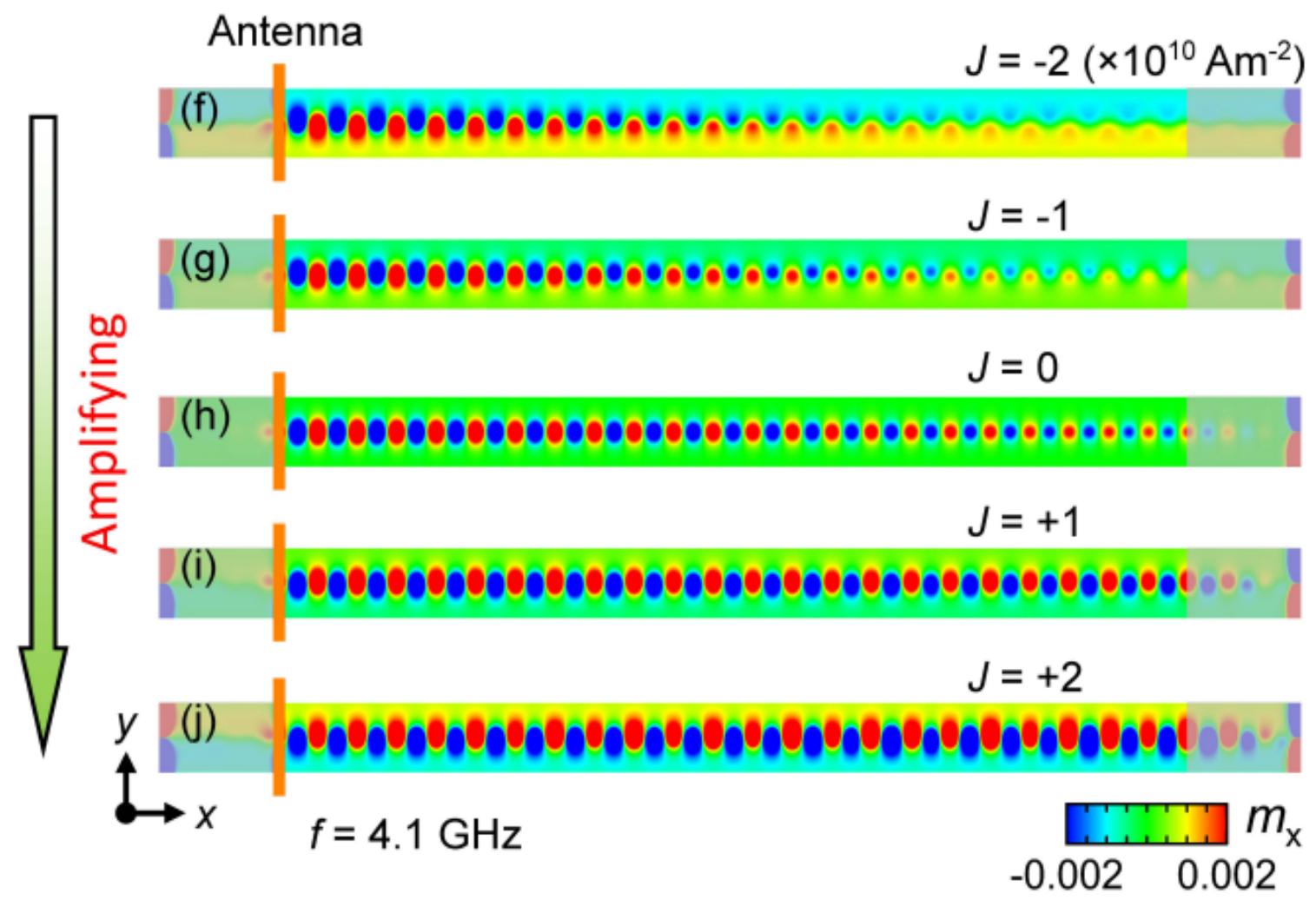

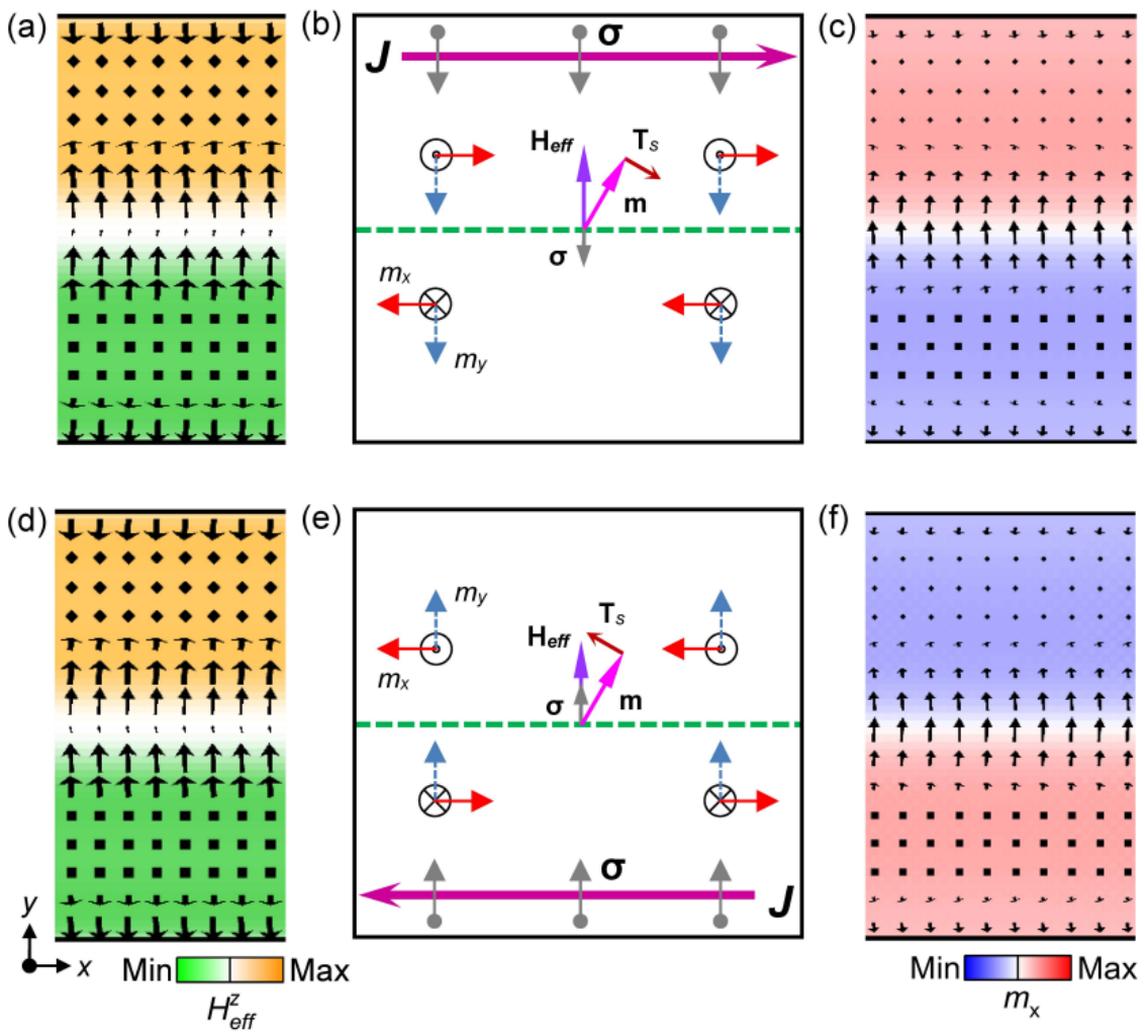

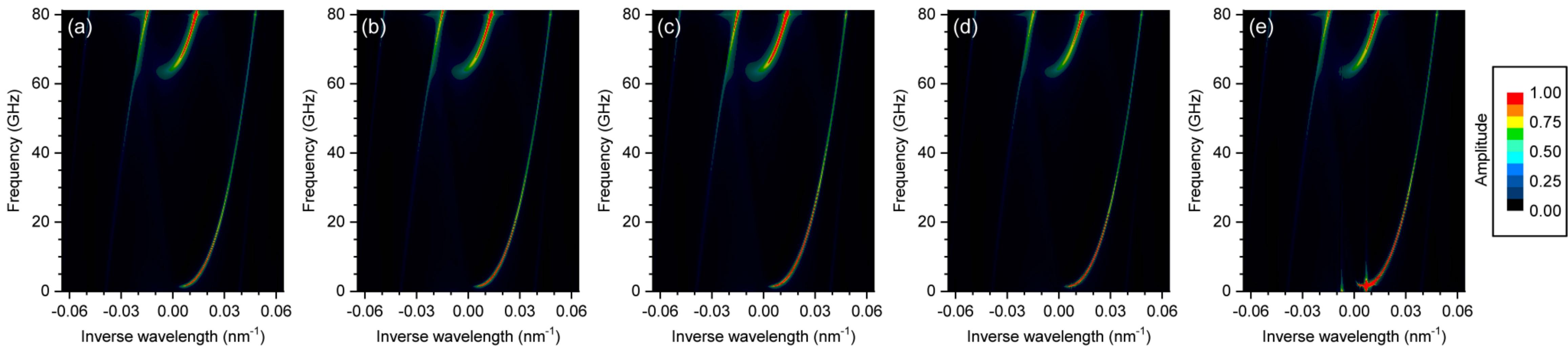

# Supplementary Material for

# Amplifying spin waves along Néel domain wall by spin-orbit torque

Xiangjun Xing,[1,a)] T. Wang,[1] and Yan Zhou[2,a)]

[1]*School of Physics & Optoelectronic Engineering, Guangdong University of Technology, Guangzhou 510006, China*

[2]*School of Science & Engineering, The Chinese University of Hong Kong, Shenzhen, Guangdong 518172, China*

**This file includes:**

SECTION I

Eqs. S1 Complex equations with respective to $\Psi = e^{-x/L}e^{i(\omega t - kx)}$.

Eqs. S2 Complex equations with respect to $m_x^0$ and $m_z^0$.

SECTION II

Fig. S1 Effect of the field-like torque on spin-wave amplification relative to the damping-like torque.

Fig. S2 Dependence of the theoretical $L(J)$ curve on the representation of $\mathbf{m}^g(y)$.

Fig. S3 Mode profile of spin waves at 5.3 GHz for various current densities.

Fig. S4 Mode profile of spin waves at 16.6 GHz for various current densities.

---
a)xjxing@gdut.edu.cn and zhouyan@cuhk.edu.cn

# SECTION I

**Equations S1** Complex equations with respective to $\Psi = e^{-x/L}e^{i(\omega t - kx)}$.

$$\gamma_0\alpha\left[-a_J m_x^{0\,2} m_z^0 - \mathcal{D}\left(-ik-\tfrac{1}{L}\right)m_x^{0\,2}m_z^0 + \mathcal{K}m_x^0 m_z^{0\,2} - a_J m_z^{0\,3} - \mathcal{D}\left(-ik-\tfrac{1}{L}\right)m_z^{0\,3}\right]\Psi^3 + \left[\tfrac{\gamma_0\alpha\mathcal{A}}{\Delta^2}m_x^0 + \gamma_0\alpha\left(-\tfrac{\mathcal{A}}{\Delta^2}+\tfrac{\mathcal{D}}{\Delta}\right)m_x^0 - \gamma_0\alpha\mathcal{A}\left(-ik-\tfrac{1}{L}\right)^2 m_x^0 - \gamma_0\alpha a_J m_z^0 - \gamma_0\alpha\mathcal{D}\left(-ik-\tfrac{1}{L}\right)m_z^0 - \gamma_0 a_J m_x^0 - \gamma_0\mathcal{D}\left(-ik-\tfrac{1}{L}\right)m_x^0 - \tfrac{\gamma_0\mathcal{A}}{\Delta^2}m_z^0 + \gamma_0\left(\tfrac{\mathcal{A}}{\Delta^2}-\tfrac{\mathcal{D}}{\Delta}\right)m_z^0 + \gamma_0\mathcal{K}m_z^0 + \gamma_0\mathcal{A}\left(-ik-\tfrac{1}{L}\right)^2 m_z^0 + i\omega m_x^0\right]\Psi = 0,$$

(S1a)

$$\left[-\tfrac{\gamma_0\alpha\mathcal{A}}{\Delta^2}m_x^{0\,2} + \gamma_0\alpha\left(\tfrac{\mathcal{A}}{\Delta^2}-\tfrac{\mathcal{D}}{\Delta}\right)m_x^{0\,2} + \gamma_0\alpha\mathcal{A}\left(-ik-\tfrac{1}{L}\right)^2 m_x^{0\,2} - \tfrac{\gamma_0\alpha\mathcal{A}}{\Delta^2}m_z^{0\,2} + \gamma_0\alpha\left(\tfrac{\mathcal{A}}{\Delta^2}-\tfrac{\mathcal{D}}{\Delta}\right)m_z^{0\,2} + \gamma_0\alpha\mathcal{K}m_z^{0\,2} + \gamma_0\alpha\mathcal{A}\left(-ik-\tfrac{1}{L}\right)^2 m_z^{0\,2} + \gamma_0 a_J m_x^{0\,2} + \gamma_0\mathcal{D}\left(-ik-\tfrac{1}{L}\right)m_x^{0\,2} - \gamma_0\mathcal{K}m_x^0 m_z^0 + \gamma_0 a_J m_z^{0\,2} + \gamma_0\mathcal{D}\left(-ik-\tfrac{1}{L}\right)m_z^{0\,2}\right]\Psi^2 = 0,$$

(S1b)

$$\gamma_0\alpha\left[a_J m_x^{0\,3} + \mathcal{D}\left(-ik-\tfrac{1}{L}\right)m_x^{0\,3} - \mathcal{K}m_x^{0\,2}m_z^0 + a_J m_x^0 m_z^{0\,2} + \mathcal{D}\left(-ik-\tfrac{1}{L}\right)m_x^0 m_z^{0\,2}\right]\Psi^3 + \left[\gamma_0\alpha a_J m_x^0 + \gamma_0\alpha\mathcal{D}\left(-ik-\tfrac{1}{L}\right)m_x^0 + \tfrac{\gamma_0\alpha\mathcal{A}}{\Delta^2}m_z^0 + \gamma_0\alpha\left(-\tfrac{\mathcal{A}}{\Delta^2}+\tfrac{\mathcal{D}}{\Delta}\right)m_z^0 - \gamma_0\alpha\mathcal{K}m_z^0 - \gamma_0\alpha\mathcal{A}\left(-ik-\tfrac{1}{L}\right)^2 m_z^0 + \tfrac{\gamma_0\mathcal{A}}{\Delta^2}m_x^0 + \gamma_0\left(-\tfrac{\mathcal{A}}{\Delta^2}+\tfrac{\mathcal{D}}{\Delta}\right)m_x^0 - \gamma_0\mathcal{A}\left(-ik-\tfrac{1}{L}\right)^2 m_x^0 - \gamma_0 a_J m_z^0 - \gamma_0\mathcal{D}\left(-ik-\tfrac{1}{L}\right)m_z^0 + i\omega m_z^0\right]\Psi = 0.$$

(S1c)

**Equations S2** Complex equations with respect to $m_x^0$ and $m_z^0$.

$$\frac{\gamma_0 \alpha \mathcal{D}}{\Delta} m_x^0 + \gamma_0 \alpha \mathcal{A} k^2 m_x^0 - \frac{\gamma_0 \alpha \mathcal{A}}{L^2} m_x^0 - \gamma_0 \alpha a_J m_z^0 + \frac{\gamma_0 \alpha \mathcal{D}}{L} m_z^0 - \gamma_0 a_J m_x^0 + \frac{\gamma_0 \mathcal{D}}{L} m_x^0 -$$

$$\frac{\gamma_0 \mathcal{D}}{\Delta} m_z^0 - \gamma_0 \mathcal{A} k^2 m_z^0 + \gamma_0 \mathcal{K} m_z^0 + \frac{\gamma_0 \mathcal{A}}{L^2} m_z^0 + i \left( -\frac{2\gamma_0 \alpha \mathcal{A} k}{L} m_x^0 + \gamma_0 \alpha \mathcal{D} k m_z^0 + \right.$$

$$\left. \gamma_0 \mathcal{D} k m_x^0 + \frac{2\gamma_0 \mathcal{A} k}{L} m_z^0 + \omega m_x^0 \right) = 0,$$

(S2a)

$$\gamma_0 \alpha a_J m_x^0 - \frac{\gamma_0 \alpha \mathcal{D}}{L} m_x^0 + \frac{\gamma_0 \alpha \mathcal{D}}{\Delta} m_z^0 + \gamma_0 \alpha \mathcal{A} k^2 m_z^0 - \gamma_0 \alpha \mathcal{K} m_z^0 - \frac{\gamma_0 \alpha \mathcal{A}}{L^2} m_z^0 + \frac{\gamma_0 \mathcal{D}}{\Delta} m_x^0 +$$

$$\gamma_0 \mathcal{A} k^2 m_x^0 - \frac{\gamma_0 \mathcal{A}}{L^2} m_x^0 - \gamma_0 a_J m_z^0 + \frac{\gamma_0 \mathcal{D}}{L} m_z^0 + i \left( -\gamma_0 \alpha \mathcal{D} k m_x^0 - \frac{2\gamma_0 \alpha \mathcal{A} k}{L} m_z^0 - \right.$$

$$\left. \frac{2\gamma_0 \mathcal{A} k}{L} m_x^0 + \gamma_0 \mathcal{D} k m_z^0 + \omega m_z^0 \right) = 0.$$

(S2b)

**SECTION II**

**Figure S1**

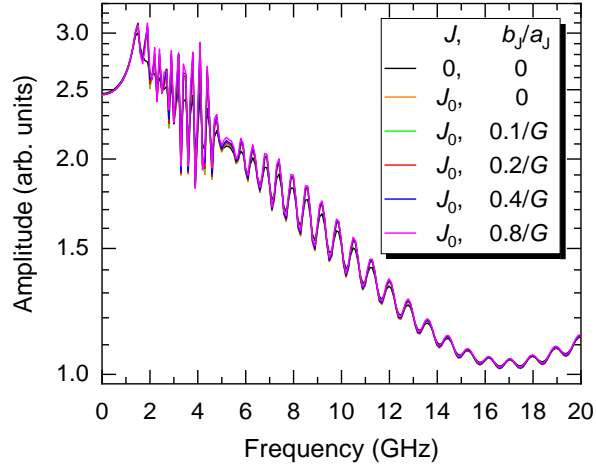

Fig. S1 Effect of the field-like torque on spin-wave amplification relative to the damping-like torque. $b_J/a_J$ represents the ratio between the field-like and damping-like torques. $J_0 = 1.0\times10^{10}$ Am$^{-2}$. $G = \frac{\Lambda^2}{(\Lambda^2+1)+(\Lambda^2-1)(\mathbf{m}\cdot\mathbf{\sigma})}$ is the spin torque asymmetry determined by the relative orientation of $\mathbf{m}$ and $\mathbf{\sigma}$ in the FM layer; in simulations $\Lambda = 2$.

Here, we consider different strengths of the field-like torque. From the comparison, it is seen that the field-like torque is almost irrelevant even for the largest $b_J/a_J$, which is unrealistically high from the experimental viewpoint.

**Figure S2**

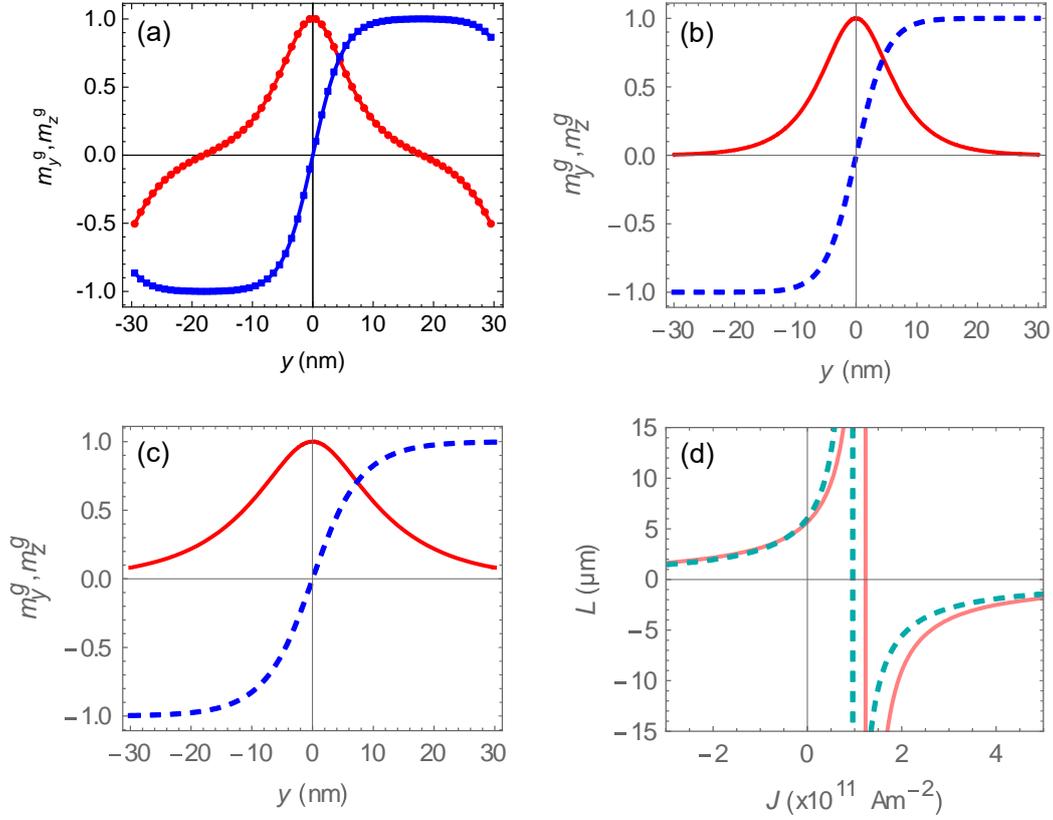

Fig. S2 Dependence of the theoretical $L(J)$ curve on the representation of $\mathbf{m}^g(y)$.

(a) Spatial profile of the ground-state magnetization configuration derived from micromagnetic simulations. A Néel domain wall is centered at $y = 0$. Circles: $m_y^g$; Squares: $m_z^g$.

(b) Spatial profile of the ansatz function A:

$\mathbf{m}^g(y) = \left(0, \mathrm{Sech}\left(\frac{y}{\Delta}\right), \mathrm{Tanh}\left(\frac{y}{\Delta}\right)\right)$. Solid line: $m_y^g$; Dashed line: $m_z^g$.

(c) Spatial profile of the ansatz function B:

$\mathbf{m}^g(y) = \left(0, \frac{2\sqrt{2}}{\pi}\sqrt{\pi \mathrm{ArcTan}\left(e^{\frac{y}{\Delta}}\right) - 2\mathrm{ArcTan}^2\left(e^{\frac{y}{\Delta}}\right)}, \frac{4}{\pi}\mathrm{ArcTan}\left(e^{\frac{y}{\Delta}}\right) - 1\right)$. Solid line: $m_y^g$; Dashed line: $m_z^g$.

(d) Theoretical $L(J)$ curves based on different forms of $\mathbf{m}^g(y)$. The solid and dashed lines correspond to the functions in (b) and (c), respectively.

This figure indicates that $L(J)$ sensitively relies on $\mathbf{m}^g(y)$.

**Figure S3**

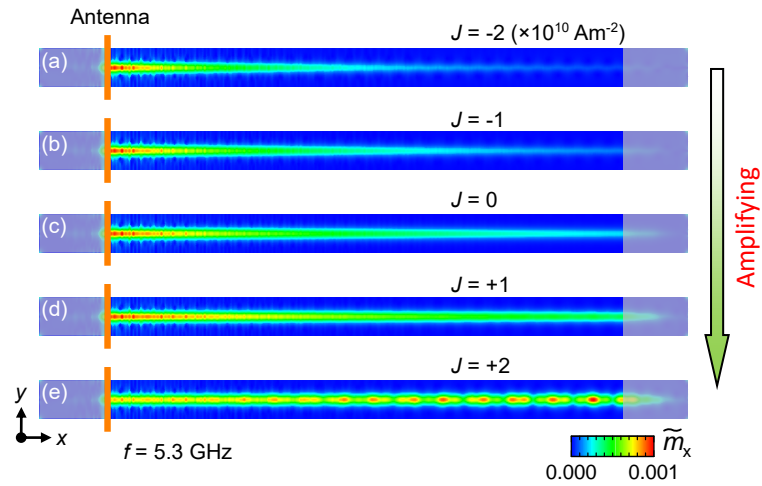

Fig. S3 Mode profile of spin waves at 5.3 GHz for various current densities. The current density is indicated in each panel of (a)–(e). With the increasing current, the amplifying effect on the spin waves manifests itself.

The oscillatory pattern to the right of the antenna, featuring a beating behavior, arises from interference of the propagating spin wave and the reflected wave from the right edge of the left shaded region. In panel (e), as the propagation distance increases, a localization feature becomes pronounced. Here, instead of the interference, the spin-wave instability (Refs. S1–S3) associated with nonlinear effect at high amplitude should play a key role.

**Figure S4**

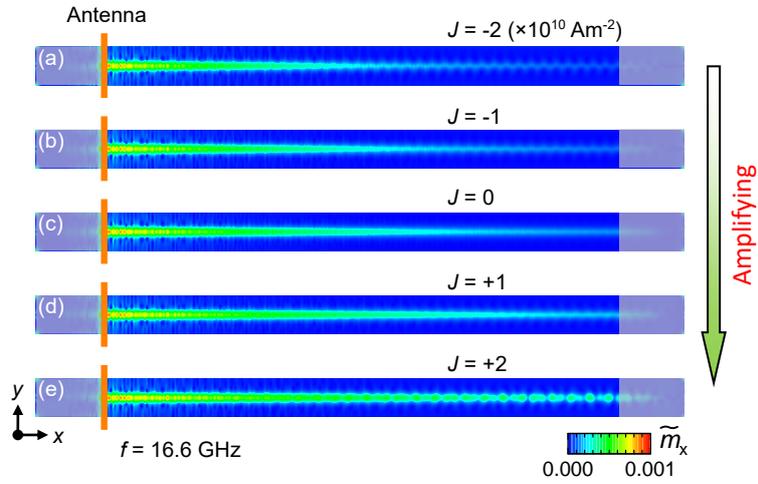

Fig. S4 Mode profile of spin waves at 16.6 GHz for various current densities. The current density is indicated in each panel of (a)–(e). With the increasing current, the amplifying effect on the spin waves manifests itself.

The oscillatory pattern to the right of the antenna, featuring a beating behavior, arises from interference of the propagating spin wave and the reflected wave from the right edge of the left shaded region. In panel (e), as the propagation distance increases, a localization feature becomes pronounced. Here, instead of the interference, the spin-wave instability (Refs. S1–S3) associated with nonlinear effect at high amplitude should play a key role.